\documentclass[10pt,letterpaper]{article}
\usepackage{setspace}
\makeatletter

\newcommand{\LyX}{L\kern-.1667em\lower.25em\hbox{Y}\kern-.125emX\@}
\newcommand{\noun}[1]{\textsc{#1}}

\newenvironment{lyxlist}[1]
  {\begin{list}{}
    {\settowidth{\labelwidth}{#1}
     \setlength{\leftmargin}{\labelwidth}
     \addtolength{\leftmargin}{\labelsep}
     }}
  {\end{list}}

\makeatother

\begin{document}

{\par\noindent \centering \textsf{\LARGE Spatial  distribution of atoms in gas-covered
Pd-X  nanoparticles (X= Ag, Cu,  Ni,  Pt)}\LARGE \par}

\bigskip{}
{\par\noindent \centering \textsf{Mahesh Menon and Badal C. Khanra{*}}\par}

{\par\noindent \centering \textsf{Condensed Matter Physics Group, Saha Institute
of Nuclear Physics}\par}

{\par\noindent \centering \textsf{1/AF, Bidhannagar, Calcutta- 700 064, India}\par}

\noindent \textsf{\textbf{\noun{Abstract}}} \textsf{}

\textsf{A  Monte-carlo (MC) simulation procedure has been developed where the
pair bond energies are allowed to take into account the various coordination
numbers of surface atoms  and the presence of adsorbates. The pair bond energies
are calculated from partial bond energies of atoms which, in turn, are calculated
from modified tight binding  model in the second moment approximation. The model
has been applied to study the role of adsorption of hydrogen, oxygen, carbon
monoxide and nitric oxide on the surface composition and surface bond geometry
of bimetallic Pd-X (X = Ag, Cu, Ni, Pt) nanoparticles having fcc cubo-octahedral
geometry with 201, 586, 1289 and 2406 atoms. The results are compared with the
known experimental results. Importance of the results in studying reactions
on supported bimetallic catalysts has been highlighted. }

\medskip{}

\noindent \textsf{{*}Author for correspondence }

\noindent \textsf{Tel.:   (91) (33) 337 5345}

\noindent \textsf{FAX: (91) (33) 337 4637 }

\noindent \textsf{e-mail:  badal@cmp.saha.ernet.in }
\newpage

\noindent \textsf{\textbf{\noun{I. INTRODUCTION}}}

\textsf{Supported bimetallic catalysts are of enormous importance in petroleum
and other chemical industries\( .^{1-3} \) They are used for better selectivity
of products in parallel reactions. Mostly, these bimetallic catalysts are supported
nanostructures, which may have specific active sites like the corner sites,
edge sites or sites on regular crystalline faces. In order to understand the
catalytic activity and selectivity of a bimetallic nanostructure  it is important
to know the average surface composition and particularly the occupancy of various
sites by A or B metal  atoms of the A-B bimetallic system. Since in catalytic
reactions the bimetallic catalyst surfaces are partially or totally covered
by reacting gases, it is also important to study the surface composition in
presence of reacting gases. It is the aim of the present work to study the spatial
distribution of atoms for a few Pd-based bimetallic nanoparticles like Pd-Ag,
Pd-Cu, Pd-Ni and Pd-Pt nanoparticles having dispersion in the range of 0.6 -
0.3 (the total number of atoms per particle being in the range of 200 - 2500).
The motivation to study Pd-based systems follows from the fact that Pd is a
very good catalyst for hydrogenation and oxidation reactions. Amongst the alloy
systems, Pd-Pt catalysts are of great technological interest because of their
use in (de)hydrogenation reactions, oxidation reaction, (dehydro)cyclization
reaction, hydrogen peroxide production etc\( ,^{4-8} \) besides their use in
electrochemical studies\( .^{9-10} \) These catalysts have the additional advantage
of being more resistant to poisoning by sulfur and nitrogen compounds than the
respective monometallic catalysts\( .^{11} \) Alloying Pd with copper and silver
(Pd-Cu, and Pd-Ag systems) is found to greatly enhance the selectivity of Pd
in the partial hydrogenation of dienes\( ^{12} \) and in the hydro-dehalogenation
of olefinic compounds\( .^{13} \) In these systems the Pd surface has to be
partly covered by the second metal (Cu or Ag) in a carefully controlled way.
This is achieved in bimetallic nanostructures. Besides, Pd-Cu systems show some
promise as catalysts for simultaneous oxidation of CO and reduction of NO\( ,^{14-15} \)
the very vital reactions for automotive pollution control. Pd-Ni systems also
show interesting features in their use as catalysts for 1,3-butadiene hydrogenation
reaction. The Pd\( _{5} \)Ni\( _{95} \) system, for example,  is found to
have greater activity than the pure metals\( .^{16} \) In view of the above
interesting features we study in this work the atomic distribution in the above
four Pd-based nanostructures in presence of different coverage (\( 0\leq \theta \leq 1 \))
of some simple adsorbates like hydrogen, oxygen, carbon monoxide (CO) and nitric
oxide (NO). The plan  of the paper is as follows: In section II we describe
the formulation of the model. In section III we apply the formalism to the four
Pd-based bimetallic nanoparticle system mentioned above. Conclusions are drawn
in section IV.}

\bigskip{}
\noindent \textsf{\textbf{\noun{II. FORMULATION OF THE MODEL}}}

\textsf{The model used in this work is based on the Monte-Carlo (MC) simulation
procedure adopted by Strohl and King\( ^{17} \) to study the segregation behavior
of Pt-Ib bimetallic nanoparticles. However, we have introduced two modifications
to their procedure as discussed below:  }

\textsf{First, the pair bond energy required for MC simulation are evaluated
taking into account the fact that the pair bond energies are dependent on the
coordination number of the atoms. Thus, the pair bond energy between two nearest
neighbor atoms,  E\( _{ij} \), is related to the site energy (or the partial
bond energy) E\( ^{^{i}}_{c}(n) \) etc. as\( ^{17} \) }

{\par\centering \textsf{\hfill{} E\( _{ij} \)  = w\( _{ij}/Z \) + E\( ^{^{i}}_{c}(n)/n \)
+ E\( ^{^{j}}_{c}(m)/m \) \hfill{} (1)}\par}

\noindent \textsf{where \( i, \) \( j \) = A or B atom of the general A\( _{X} \)B\( _{X-1} \)
alloy and w\( _{ij} \) = 0 if \( i \) \( = \) \( j \). E\( ^{^{i}}_{c}(n) \)
is the  partial bond energy of the \( i \)-atom with \( n \) nearest neighbors
etc. w\( _{ij} \) is the interchange energy for two dissimilar atoms which
may be estimated from the excess heat of mixing; and \( Z \) is the bulk coordination
number. In order to find E\( ^{^{i}}_{c}(n) \) Strohl and King used an empirical
formula}

\noindent \textsf{\hfill{} E\( ^{^{i}}_{c}(n) \) = \( a^{^{i}} \) + \( b^{^{i}}n \)
+ \( c^{^{i}}n^{^{2}} \) \hfill{} (2)}

\noindent \textsf{where \( a^{^{i}} \), \( b^{^{i}} \) and \( c^{^{i}} \)
are constants for the \( i \)-metal and may be obtained from the experimental
values of the heat of sublimation, the energy of bulk vacancy formation and
the surface energy for (100) surface.}

\textsf{We use the \( a \), \( b \) and \( c \) values evaluated by Rousset
et al.\( ^{18} \) from a modified tight binding method (MTB).  Here, the parameters
are determined from the experimental values of dimer energy, surface energy
and the energy for monovacancy creation. In this MTB scheme the site energy
for six-coordinated corner atom and the seven-coordinated edge atoms can be
obtained from interpolation rather than extrapolation as was obtained  in the
work of Strohl and King. Once the site energies are known, the pair bond energies
can be easily evaluated\( .^{19} \) The second modification is made to consider
the chemisorption. For the present study we have introduced necessary changes
in the MC simulation to take into account the effect of chemisorption. Here
we take an additional term \( \theta  \)(E\( _{A} \)-E\( _{B} \)) in the
configuration energy where \( \theta  \) is the adsorbate coverage and E\( _{A(B)} \)
is the chemisorption energy of the adsorbate on \( A(B) \) metal\( .^{20-21} \)
Once the constants \( a^{^{i}} \), \( b^{^{i}} \) and \( c^{^{i}} \) are
derived and the parameters w\( _{ij}/Z \), \( \theta  \) and (E\( _{A} \)-E\( _{B} \))
are provided, the MC simulation is carried out in the usual manner until the
equilibrium configuration energy is found. In essence, the atomic distribution
results obtained in our nanostructure work would complement the segregation
results obtained earlier by Rousset et al.\( ^{18} \) for the semi-infinite
Pd-X alloy systems. In addition, the present results consider the role of various
degrees of coverage of adsorbates on the atomic distribution.  }

\bigskip{}
\noindent \textsf{\textbf{\noun{III. APPLICATIONS }}}

\textsf{In view of the fact that all the constituent metals of the bimetallic
systems considered in this work have fcc structure, the particles with number
of atoms in the range of 200 - 2500 may be assumed to have the fcc cubo-octahedral
geometry. With this assumption we perform the MC simulation. The important input
parameters used in the work are presented in Tables 1 and 2. In this simulation
program the important variable parameters are (a) the bimetallic system, (b)
the bulk composition, (c) the adsorbate coverage, and (d) the temperature. Since
the composition variation with temperature is well-known, we have performed
the simulation for a typical alloy equilibration temperature (700K). Though
we have performed simulation for several bulk compositions (20\% Pd, 50\% Pd,
66\% Pd, 80\% Pd), for convenience of comparison of several systems in presence
of different adsorbates with varying coverage we would present results mostly
for Pd\( _{50} \)X\( _{50} \) composition. The simulation results being very
exhaustive cannot all be presented in this paper. We would be presenting some
typical results of the systems. Readers interested in the details for other
compositions, particle sizes, coverage or adsorbates may kindly contact the
authors.   }

\bigskip{}
\noindent \textsf{\textbf{A. Average surface Pd concentration  }}

\textsf{In figure 1(a) we have plotted for the Pd\( _{50} \)Ag\( _{50} \)
nanosystem the fraction of surface sites occupied by Pd atoms for different
particle size and H coverage. In Figs. 1(b), (c) and (d) similar results are
plotted for the oxygen, CO and NO adsorption. Several characteristics are to
be noted: }

\begin{description}
\item [\textsf{(i)}]\textsf{For clean Pd\( _{50} \)Ag\( _{50} \) nanoparticles of
all sizes Ag segregates to the surface sites.}
\item [\textsf{(ii)}]\textsf{As the coverage increases the fraction of Pd surface
atom increases leading to a reduction in Ag segregation and ultimately Pd atoms
tend to segregate to the surface sites.}
\end{description}
\textsf{This is true for all particle sizes and all the four adsorbates under
consideration. In case of CO adsorbate Pd atoms seem to segregate to the surface
sites even at coverage 0.25, while for other gases the segregation reversal
takes place at coverage close to one. This behavior can be interpreted as a
result of the higher heat of adsorption of adsorbates on Pd compared to that
on Ag. Higher binding energy of adsorbates causes Pd atoms to move to the surface
countering the normal segregating force. The difference in heat of CO adsorption
on Pd and Ag being very large (\( \approx  \)28 kcals/mole), Pd atoms start
segregating to the surface even at low coverage. It may be noticed that for
clean and small coverage of adsorbates the Ag surface fraction increases as
the particle size increases (smaller dispersion). This follows from the fact
that the fraction of surface sites decreases with increase in particle size;
and therefore, the Ag atoms, if available, can cover higher fraction of surface
sites for clean and small coverage (please see Table 4 of Strohl and King\( ^{17} \)).
However, as the coverage increases, more Pd atoms move to the surface due to
stronger binding to the adsorbate; and thus, the Pd surface fraction increases
with increase in particle size. }

\textsf{Figs. 2(a)-(d) show the Pd surface fraction for the Pd\( _{50} \)Cu\( _{50} \)
nanoparticles in presence of the adsorbed gases. For this bimetallic system,
Cu segregates to the surface for hydrogen coverage up to 0.6 for larger particles.
For smaller particles Cu segregates for hydrogen coverage even up to 0.75. For
still larger coverage Pd segregates to the surface. For oxygen as adsorbates
on Pd\( _{50} \)Cu\( _{50} \) system, Cu segregates for all coverages and
all particle sizes studied in this work. Higher the coverage, higher is the
extent of Cu  segregation. This again follows from the heat of adsorption difference
criterion mentioned before. Here binding energy of oxygen on copper is higher
than that on Pd. For CO and NO adsorption on the Pd\( _{50} \)Cu\( _{50} \)
nanoparticles Cu segregates for clean and very low coverage. For CO and NO coverage
\( > \)0.25 the Pd atoms start segregating to the surface.}

\textsf{The difference in the heat of  adsorption of hydrogen (oxygen) on pure
Pd and Ni metals is very small. This causes only small changes in the surface
Pd concentration with coverage for the Pd\( _{50} \)Ni\( _{50} \) system (Figs.
3(a) and (b)). For all particle sizes and coverage up to one Ni segregates to
the surface. For CO and NO adsorption on Pd\( _{50} \)Ni\( _{50} \), on the
other hand, Pd atoms start segregating to the surface from coverage 0.2 (please
see Figs. 3(c) and (d)). The segregation behavior of the Pd\( _{50} \)Pt\( _{50} \)
system is very simple. The clean system shows Pd segregation. And for all particle
sizes and all adsorbates Pd surface concentration slowly increases with increase
in coverage (please see Figs. 4(a)-(d)).}

\textsf{In order to compare the segregation behavior of the bimetallic Pd\( _{50} \)X\( _{50} \)
nanosystems in presence of different adsorbates we plot in figure 5(a)-(d) the
Pd surface fraction for a particular coverage, \( \theta  \) = 0.5. It may
be noticed that in presence of hydrogen or oxygen (Figs. 5(a) and (b) respectively)
Pd segregates to the surface sites only for the Pd-Pt system. For the rest of
the systems the other component segregates. In presence of CO, however, Pd segregates
for all systems and for all particle sizes (Fig. 5(c)). In presence of NO, only
for the Pd-Ag system Ag segregates, while for the rest of the systems Pd segregates
to the surface sites (Fig.5(d)). The role of adsorbate coverage on the surface
Pd concentration for the different bimetallic systems has been shown in Figs.
6(a)-(d) for the four different adsorbates. The results are for particles having
586 atoms (dispersion D = 0.457). The figures are self-explanatory. Except for
hydrogen on Pd\( _{50} \)Ni\( _{50} \) (Fig. 6(a)) and oxygen on Pd\( _{50} \)Ni\( _{50} \)
and Pd\( _{50} \)Cu\( _{50} \) (Fig. 6 (b)) for all other systems Pd surface
fraction increases with coverage. }

\bigskip{}
\noindent \textsf{\textbf{B. Site occupancy and surface coordination}}

\textsf{Besides the overall surface Pd concentration for Pd-X nanoparticles
it is also very important to know the occupancy of various types of surface
sites on the fcc cubo-octahedron particle. On the particle surfaces there are
6-coordinated corner sites, 7-coordinated edge sites, 8-coordinated fcc(100)-like
face sites and 9-coordinated  fcc(111)-like face sites. In figures 7(a)-(d)
we have shown how for CO adsorption on 586-atom nanoparticles of the four bimetallic
systems the fraction of Pd atoms on the 6, 7, 8 and 9-coordinated surface sites
changes with coverage. For CO on Pd\( _{50} \)Ag\( _{50} \) system (Fig. 7(a))
at low coverage the 6 and 7 -coordinated sites are generally occupied by Ag
atoms. As the coverage increases, gradually all types of sites get increasingly
occupied by Pd atoms. And at full monolayer coverage except the 8 -coordinated
sites all other sites are occupied by Pd atoms. This is also the case for Pd\( _{50} \)Cu\( _{50} \)
(Fig. 7(b)) system. In case of Pd\( _{50} \)Ni\( _{50} \) system (Fig. 7(c))
no particular type of surface sites is occupied fully by Pd or Ni atoms at any
of the coverage from zero to one. In case of Pd\( _{50} \)Pt\( _{50} \) system
the occupancy of the 9-coordinated sites changes very slowly from 0.72 to 0.78
as the coverage increases from zero to one. For NO adsorption the coverage-dependent
site occupancy results for the four bimetallic systems are shown in figures
8(a) to (d). The results are self-explanatory.}

\textsf{For studying the relative occupancy of various sites in the four bimetallic
systems we plot in figures 9(a)-(b) the site occupancy in the systems for a
particular coverage (\( \theta  \)) and a particular particle size (586-atom
nanoparticles). The adsorbates considered are CO and NO. For both the adsorbates,
it may be noticed that at this 0.5 coverage the 6-coordinated sites are almost
fully occupied by Pd in case of Pd\( _{50} \)Pt\( _{50} \) system. On the
other hand, for the Pd\( _{50} \)Ag\( _{50} \) and Pd\( _{50} \)Cu\( _{50} \)
systems the 6 -coordinated sites are generally occupied by Ag and Cu atoms respectively.}

\bigskip{}
\noindent \textsf{\textbf{C. Size dependence of surface site occupancy}} \textsf{}

\textsf{Particle size, quite often, plays an important role in catalysis. In
bimetallic systems, this has an additional significance. Because, different
metallic atoms may preferentially occupy different type of sites. In order to
study this aspect we show in figure 10(a)-(d) how for a CO coverage of 0.5 the
surface sites of a 586-atom nanoparticle are preferentially occupied by one
type of atom or the other. What we find is that there is no general rule for
occupancy of various sites as the particle size increases as there are crossovers
of curves for different particle sizes in each of the four systems. In case
of Pd\( _{50} \)Ag\( _{50} \) system (Fig. 10(a)) the occupancy of 8 and 9-coordinated
sites by Pd atoms does not change much with particle size. But there are fluctuations
in the occupancy of 6 and 7-coordinated sites by Pd atoms as the particle size
increases. Similarly, in case of Pd\( _{50} \)Cu\( _{50} \) and Pd\( _{50} \)Ni\( _{50} \)
system there are fluctuations in occupancy of all types of surface sites as
the particle size increases (Figs. 10(b) and (c)). In case of Pd\( _{50} \)Pt\( _{50} \)
we find some systematic variation in the occupancy of various surface sites.
As the number of atoms in the particle increases from 201 to 2406, the occupancy
of 9-coordinated sites by Pd atoms increases significantly. This variation is
distinctively different from other bimetallic systems. }

\textsf{The net result of the model presented in this work is that we have simulated
the probable surface composition and the atomic distribution for the nanoparticles
of some Pd-based bimetallics in presence of various coverages of simple gases
like hydrogen, oxygen, CO and NO. However, the experimental results are scarce,
particularly for adsorbed systems. In the zero-coverage limit, i.e. for clean
bimetallic nanoparticles experimental results exist for the PdPt and PdCu systems.
The model shows Pd segregation for the PdPt system and Cu segregation for the
Pd-Cu system in agreement with the experimental results\( .^{19,26} \) But
the coverage-dependent surface composition and site occupancy are yet to be
experimentally realized. However, in the absence of experimental results, the
present model can serve some purpose in estimating the probable surface composition
for the nanoparticles under various coverage of gases. Besides, the knowledge
of the site occupancy by A type atom or by B type atom of A-B bimetallic atoms
may sometime help to identify the active sites on a catalyst for a particular
catalytic reaction. For example, it has been found that the rate of 1,3-butadiene
hydrogenation reaction on a Pd\( _{5} \)Ni\( _{95} \) system is larger than
that on pure metals\( .^{16} \) This may be due to the ensemble effect whereby
a particular form of ensemble of Pd and Ni atoms form the active site. Such
behavior may be explained from a knowledge of the site occupancy as studied
in this work\( .^{27} \)  }

\bigskip{}
\noindent \textsf{\textbf{\noun{IV. CONCLUSIONS}}}

\textsf{Monte-carlo simulations are carried out to find out the nature of occupancy
of surface sites of Pd-based bimetallic nanoparticles. We upgraded the earlier
MC simulation scheme of Strohl and King\( ^{17} \) with input data obtained
from improved cohesive energy calculation\( .^{18} \) }

\begin{lyxlist}{00.00.0000}
\item [\textsf{\hfill{}(a)}]\textsf{It has shown how the overall surface composition
varies with particle size, the nature of adsorbed gas and its  coverage . }
\item [\textsf{\hfill{}(b)}]\textsf{It has also been shown how for a particular adsorbed
gas the surface composition varies from one bimetallic system to the other.}
\item [\textsf{\hfill{}(c)}]\textsf{We have also shown how different types of surface
sites of the bimetallic systems are occupied and how the occupancy of these
sites varies with gas coverage.}
\item [\textsf{\hfill{}(d)}]\textsf{We have discussed briefly the importance of such
studies in understanding the role of active sites in catalysis.}
\end{lyxlist}
\noindent \textsf{\hfill{} The validity of the model would be justified  after
 more and more experimental data and particularly for gas-covered surfaces,
are available. }

\bigskip{}
\noindent \textsf{\textbf{ACKNOWLEDGMENTS:}} \textsf{}

\textsf{Mahesh Menon thanks the Council of Scientific and Industrial Research,
New Delhi, for the award of a Senior Research Fellowship (No. 9/420/(12)/95-EMR-I)
to carry out this work. BCK acknowledges receipt of the basic MC program for
segregation studies from Prof. Terry King.  }

\bigskip{}
\noindent \textsf{\textbf{REFERENCES}} \textsf{ }

\begin{enumerate}
\item \textsf{J. H. Sinfelt,} \textsf{\emph{Bimetallic Catalysts}} \textsf{(Wiley,
New York, 1983).}
\item \textsf{V. Ponec,  Adv. Catal.} \textsf{\textbf{32}}\textsf{, 149 (1983). }
\item \textsf{V. Ponec, Catal. Rev. Sci. Eng.} \textsf{\textbf{18}}\textsf{, 151 (1978). }
\item \textsf{L. C. A. Van den Oetelaar, O. W. Nooij, S, Oerlemans, A. W. Denier van
der Gon, H. H. Brongersma, L. Lefferts, A. G. Roosenbrand,  and  J. A. R. Van
Veen, J. Phys. Chem.} \textsf{\textbf{B 102}}\textsf{, 3445 (1998).}
\item \textsf{D. W. McKee, and P. J. Norton, J. Catal.} \textsf{\textbf{3}}\textsf{,
252 (1964). }
\item \textsf{Y. Deng, and L. An, Appl. Catal.} \textsf{\textbf{A 119}}\textsf{, 13
(1994).}
\item \textsf{S. Vigneron, P. Deprelle, and J. Hermia, Catal. Today} \textsf{\textbf{27}}\textsf{,
229 (1996).}
\item \textsf{T. Koscielski, Z. Karpinski, and Z. Paal, J. Catal.} \textsf{\textbf{77}}\textsf{,
539 (1982).}
\item \textsf{K. V. Ramesh, P. R. Sarode, S. Vasudevan, and A. K. Shukla, J. Electroanal.
Chem} \textsf{\textbf{223}}\textsf{, 91 (1987).}
\item \textsf{G. A. Attard, and R. Price, Surf. Sci.} \textsf{\textbf{335}}\textsf{,
63 (1995). }
\item \textsf{T. B. Lin, C. A. Jan, and J. R. Chang, Ind. Eng. Chem. Res.} \textsf{\textbf{34}}\textsf{,
4284 (1995).}
\item \textsf{J. Philips, A. Auroux, G. Bergeret, J. Massardier, and A. Renouprez,
J. Phys. Chem.} \textsf{\textbf{97}}\textsf{, 3565 (1993).}
\item \textsf{E. N. Balko, E. Przybylski, and F. Von Trentini, Appl. Catal.} \textsf{\textbf{B2}}\textsf{,
1 (1993). }
\item \textsf{Y. Debauge, P. Ruiz, J. Massardier, M. Abon, and J. C. Bertolini, Proceedings
of 3\( ^{rd} \) International Conference on Combustion Technologies for a clean
environment (Lisbon, Portugal, 1995) Vol. 2, Paper No. 29.1.}
\item \textsf{Y. Debauge, M. Abon, J. C. Bertolini, J. Massardier, and A. Rochefort,
Appl. Surf. Sci.} \textsf{\textbf{90}}\textsf{, 15 (1995). }
\item \textsf{P. Miegge, J. L. Rousset, B. Tardy, J. Massardier, and J. C. Bertolini,
J. Catal.} \textsf{\textbf{149}}\textsf{, 404 (1994).}
\item \textsf{J. K. Strohl, and T. S. King, J. Catal.} \textsf{\textbf{116}}\textsf{,
540 (1989).}
\item \textsf{J. L. Rousset, J. C. Bertolini ,and P. Miegge, Phys. Rev.} \textsf{\textbf{B
53}}\textsf{, 4947 (1996).  }
\item \textsf{J. L. Rousset, B. C. Khanra, A. M. Cadrot, F. Cadete Santos Aires, A.
Renouprez, and M. Pellarin, Surf. Sci.} \textsf{\textbf{352-354}}\textsf{, 583
(1996).}
\item \textsf{T. S. King, and R. G. Donnelly, Surf. Sci.} \textsf{\textbf{141}}\textsf{,
417 (1984). }
\item \textsf{B. C. Khanra and M. Menon, Int. J. Mod. Phys. B (in press). }
\item \textsf{G. Ertl, in} \textsf{\emph{The Nature of the Surface Chemical Bond}}\textsf{,
edited by T. N. Rhodin and G. Ertl (North-Holland, Amsterdam, 1979) p.319. }
\item \textsf{E. Shustorovich, Surf. Sci. Rep.} \textsf{\textbf{6}}\textsf{, 1 (1986). }
\item \textsf{D. Tomanek, S. Mukherjee, V. Kumar, and K. H. Bennemann, Surf. Sci.}
\textsf{\textbf{114}}\textsf{, 11 (1982). }
\item \textsf{J. L. Gland, B. A. Sexton, and G. Fisher, Surf. Sci.} \textsf{\textbf{95}}\textsf{,
587 (1980). }
\item \textsf{A. J. Renouprez, K. Lebas, G. Bergeret, J. L. Rousset, and P. Delichere,
        Studies in Surface Science and Catalysis} \textsf{\textbf{101}}\textsf{,
1105 (1996). }
\item \textsf{B. C. Khanra and M. Menon, Chem. Phys. Lett. (in Press)}
\end{enumerate}
\newpage
\textsf{\textbf{FIGURE CAPTIONS}}

\begin{lyxlist}{00.00.0000}
\item [\textsf{Fig.~1.}]\textsf{MC simulated surface fraction of Pd atoms on bimetallic
Pd\( _{50} \)Ag\( _{50} \) nanoparticles as a function of dispersion and coverage
of various adsorbates: (a) Adsorbate: hydrogen; (b) Adsorbate: oxygen; (c) Adsorbate:
CO; (d) Adsorbate: NO.}
\item [\textsf{Fig.~2.}]\textsf{MC simulated surface fraction of Pd atoms on bimetallic
Pd\( _{50} \)Cu\( _{50} \) nanoparticles as a function of dispersion and coverage
of various adsorbates: (a) Adsorbate: hydrogen; (b) Adsorbate: oxygen; (c) Adsorbate:
CO; (d) Adsorbate: NO.}
\item [\textsf{Fig.~3.}]\textsf{MC simulated surface fraction of Pd atoms on bimetallic
Pd\( _{50} \)Ni\( _{50} \) nanoparticles as a function of dispersion and coverage
of various adsorbates: (a) Adsorbate: hydrogen; (b) Adsorbate: oxygen; (c) Adsorbate:
CO; (d) Adsorbate: NO.}
\item [\textsf{Fig.~4.}]\textsf{MC simulated surface fraction of Pd atoms on bimetallic
Pd\( _{50} \)Pt\( _{50} \) nanoparticles as a function of dispersion and coverage
of various adsorbates: (a) Adsorbate: hydrogen; (b) Adsorbate: oxygen; (c) Adsorbate:
CO; (d) Adsorbate: NO.}
\item [\textsf{Fig.~5.}]\textsf{MC simulated surface fraction of Pd atoms on bimetallic
Pd\( _{50} \)X\( _{50} \) nanoparticles as a function of dispersion and 0.5
coverage of various adsorbates. (X = Ag, Cu, Ni, Pt). (a) Adsorbate: hydrogen;
(b) Adsorbate: oxygen; (c) Adsorbate: CO; (d) Adsorbate: NO.}
\item [\textsf{Fig.~6.}]\textsf{MC simulated surface fraction of Pd atoms on 586-atom
bimetallic Pd\( _{50} \)X\( _{50} \) nanoparticles as a function of coverage
of various adsorbates. (X = Ag, Cu, Ni, Pt).  (a) Adsorbate: hydrogen; (b) Adsorbate:
oxygen; (c) Adsorbate: CO; (d) Adsorbate: NO.}
\item [\textsf{Fig.~7.}]\textsf{MC simulated surface fraction of Pd atoms at N-coordinated
surface sites (N= 6-9) of a 586-atom Pd\( _{50} \)X\( _{50} \) nanoparticle
in presence of various coverage of CO. (a) X= Ag; (b) X= Cu; (c) X=Ni; (d) X=
Pt.}
\item [\textsf{Fig.~8.}]\textsf{MC simulated surface fraction of Pd atoms at N-coordinated
surface sites (N= 6-9) of a 586-atom Pd\( _{50} \)X\( _{50} \) nanoparticle
in presence of various coverage of NO. (a) X= Ag; (b) X= Cu; (c) X=Ni; (d) X=
Pt.}
\item [\textsf{Fig.~9.}]\textsf{MC simulated surface fraction of Pd atoms at N-coordinated
surface sites (N= 6-9) of a 586-atom Pd\( _{50} \)X\( _{50} \) nanoparticle
(X = Ag, Cu, Ni, Pt) for 0.5 coverage of adsrbates. (a) Adsorbate: CO; (b) Adsorbate:NO.}
\item [\textsf{Fig.~10.}]\textsf{MC simulated surface fraction of Pd atoms at N-coordinated
surface sites (N= 6-9) of Pd\( _{50} \)X\( _{50} \) nanoparticle of different
sizes in presence of 0.5 coverage of CO. (a) X= Ag; (b) X= Cu; (c) X= Ni; (d)
X= Pt.}
\end{lyxlist}
\newpage
{\par\noindent \centering \textsf{Table 1}\par}

{\par\noindent \centering \textsf{Input energy parameters for Pd-X bimetallic
particles }\par}

\vspace{0.57cm}
{\centering \begin{tabular}{|c|c|c|c|c|c|}
\hline 
system&
w\textsf{\( _{Pd-X} \)}&
\textsf{E\( _{H} \)}&
\textsf{E\( _{O} \)}&
\textsf{E\( _{CO} \)}&
\textsf{E\( _{NO} \)}\\
\hline 
\hline 
Pd-X&
(ev)&
(kcals/mol)&
(kcals/mol)&
(kcals/mol)&
(kcals/mol)\\
\hline 
\hline 
Cu&
-0.0197&
-56\textsf{\( ^{^{a}} \)}&
-103\textsf{\( ^{^{a}} \)}&
-12\textsf{\( ^{^{a}} \)}&
-14\textsf{\( ^{^{e}} \)}\\
\hline 
Ag&
-0.00957&
-56\textsf{\( ^{^{a}} \)}&
-80\textsf{\( ^{^{a}} \)}&
-6\textsf{\( ^{^{a}} \)}&
-25\textsf{\( ^{^{a}} \)}\\
\hline 
Ni&
-0.0095&
-63\textsf{\( ^{^{a}} \)}&
-90\textsf{\( ^{^{c}} \)}&
-27\textsf{\( ^{^{a}} \)}&
-25\textsf{\( ^{^{a}} \)}\\
\hline 
Pt&
-0.00396&
-61\textsf{\( ^{^{b}} \)}&
-85\textsf{\( ^{^{d}} \)}&
-32\textsf{\( ^{^{a}} \)}&
-27\textsf{\( ^{^{a}} \)}\\
\hline 
Pd&
0&
-62\textsf{\( ^{^{a}} \)}&
-87\textsf{\( ^{^{a}} \)}&
-34\textsf{\( ^{^{a}} \)}&
-31\textsf{\( ^{^{a}} \)}\\
\hline 
\end{tabular}\par}
\vspace{0.57cm}

{\par\centering \textsf{\( ^{(a)} \) Ref. 22, \( ^{(b)} \) Ref. 23, \( ^{(c)} \)
Ref. 24, \( ^{(d)} \) Ref. 25, \( ^{(e)} \) Ref. 15}\par}
\bigskip{}

\bigskip{}
{\par\centering \textsf{Table 2}\par}

{\par\centering \textsf{a, b, and c parameters for different metals}\par}

\vspace{0.57cm}
{\centering \begin{tabular}{|c|c|c|c|c|c|}
\hline 
&
Pd&
Cu&
Ag&
Ni&
Pt\\
\hline 
\hline 
a&
-0.17702 &
-0.34040 &
-0.25866&
-0.42578 &
-0.42874 \\
\hline 
b&
- 0.04842 &
-0.01177 &
-0.013283 &
-0.01447 &
-0.04750 \\
\hline 
c&
0.00299&
0.00129&
0.00119&
0.00160&
0.00355\\
\hline 
\end{tabular}\par}
\vspace{0.57cm}

\end{document}